\documentclass[epj]{svjour}
\usepackage{graphics}
\usepackage{feynmf}
\usepackage[numbers]{natbib}
\usepackage{color}
\usepackage[english]{babel} 
 \def\sbar{\overline}
\def\stilde{\widetilde}
\begin{document}
\title{SUSY  dark matter(s)}
\author{Riccardo Catena \and Laura Covi
}                     
%
%
\institute{Institut f\"ur Theoretische Physik, Friedrich-Hund-Platz 1, D-37077 G\"ottingen, Germany}
\date{Received: date / Revised version: date}
%
\abstract{
We review here the status of different dark matter candidates 
in the context of supersymmetric models, in particular the neutralino 
as a realization of the WIMP-mechanism and the gravitino. 
We give a summary of the recent bounds in direct and indirect detection
and also of the LHC searches relevant for the dark matter question.
We discuss also the implications of the Higgs discovery for the 
supersymmetric dark matter models and give the prospects for the future years.
} 
\maketitle
\section{Introduction}
\label{intro}
One among the most compelling evidences for physics beyond the Standard Model (SM) of particle physics 
is the presence of an unidentified dark matter (DM) component in the observed Universe. 
This ``dark matter problem'' consists in the lack of a microscopic theory for the in\-vi\-si\-ble 
form of matter which determines the motion of stars and galaxies in many astronomical systems, supports the 
large scale cosmological structure formation and constitutes about 80\% of the total matter in the Universe~\cite{Bergstrom:2000pn,Bertone:2004pz}. 
Postulated to explain the high velocity dispersion of gala\-xies in a nearby galaxy cluster in 1933 by Zwicky, the DM hypothesis is nowadays 
corroborated by a plethora of complementary cosmological and astrophysical observations. Surveys performed from the largest structures 
we see in the Universe, namely galaxy clusters, down to dwarf and low surface brightness galaxies provide incontro\-ver\-ti\-ble evidence in favor 
of the existence of a DM component in the Universe. Within the picture one obtains combining this variety of information, DM behaves like 
a dissipation-less and collision-less fluid manifesting itself only gravitationally. The microscopic nature of the DM component of the Universe remains however unknown.

One of the most popular attempts to solve the DM problem is the celebrated paradigm where the DM candidate is a 
weakly interacting massive particle (WIMP). In this scenario DM is made of a beyond-the-Standard-Model (BSM) particle, which is stable, 
initially in thermal equilibrium in the early Universe and decoupling as a non-relativistic species. The present cosmological density for such 
a state scales approximately with the inverse of its pair annihilation rate into lighter SM particles and it can be accurately calculated by solving 
the Boltzmann equation for the WIMP number density $n_\chi$~\cite{Gondolo:1990dk,Edsjo:1997bg}:
\begin{equation} 
\label{eq:Boltzmann}
  \frac{dn_\chi}{dt} =
  -3Hn_\chi - \langle \sigma_{\rm{eff}} v \rangle
  \left( n_\chi^2 - n_{\chi,{\rm{eq}}}^2 \right)
\end{equation}
where the effective thermally averaged annihilation cross section, $\langle \sigma_{\rm{eff}} v \rangle$, accounts for DM annihilations and coannihilations. 
Its expression in terms of particle couplings and the details of the thermal average can be found in Refs.~\cite{Gondolo:1990dk,Edsjo:1997bg}. 
Qualitatively, Eq.~(\ref{eq:Boltzmann}) tells us that the present WIMP number density is determined by the competition of two phenomena leading 
to a departure from the WIMP equilibrium number density $n_{\chi,\rm{eq}}$: the expansion of the Universe, which occurs with the rate $H$, 
and the DM annihilations, characterized by the rate $\Gamma_{ann} = n_\chi \langle \sigma_{\rm{eff}} v \rangle$. The present value of $n_{\chi}$ is 
independent from the initial conditions, whose memory is erased in the thermal equilibrium phase. In practice, Eq.~(\ref{eq:Boltzmann}) has to 
be solved numerically and the very precise determination of the relic density achieved by current CMB experiments turns out to be a very tight 
constraint for many of the WIMP models proposed in the literature. 
The popularity of this framework relies on its very rich phenomenology and its straightforward implementation in many BSM models.

Another equally possible way to produce DM is instead to relax the assumption of equilibrium and consider the thermal evolution of 
the particle's abundance in the primordial plasma from a particular initial density, often taken to be zero after the inflationary dilution. 
In such a case, depending on the couplings and mass of the DM candidate, different annihilation or decay rates into such particles
allow to match the present DM energy density. For this mechanism, usually candidates interacting very weakly are favored since it 
is vital that the particles do not reach thermal equilibrium: here we will discuss in fact particles with non-renormalizable couplings 
suppressed by a large mass scale, like the Planck mass $ M_P$ for the gra\-vi\-tino or the Peccei-Quinn scale for the axino.
The thermal-plasma-contribution to their present abundance is often proportional to the maximal thermal bath temperature, since 
the non-renormalizable couplings are more effective at high temperature~\cite{Bolz:2000fu, Pradler:2006qh,Rychkov:2007uq, 
Covi:2001nw, Brandenburg:2004du, Strumia:2010aa}.
Otherwise also thermal bath particle decays in equilibrium can give a sufficiently large contribution to the DM density in what is
called the ``FIMP'' mechanism \cite{McDonald:2001vt,Covi:2002vw,Asaka:2005cn,Hall:2009bx,Cheung:2011nn},  
which is instead independent of the initial conditions and temperature.
Within this framework there is also the possibility to generate a DM population still exploiting the WIMP mechanism, via the decay of 
the decoupled WIMP into the lighter and more weakly interacting state. Such a scenario has been called 
``SuperWIMP" mechanism \cite{Covi:1999ty, Feng:2003xh, Feng:2003uy}.
 
Supersymmetry (SUSY) is one of the most popular extensions of the SM and is flexible enough to offer DM candidates of both types, as we will discuss in
detail in the following. After a brief introduction to the basics of SUSY, presented in section \ref{SUSYnut}, we review first the most popular 
SUSY dark matter candidates, in section \ref{candidates}, and then the strategies designed to detect them, in section \ref{detection}. 
Section \ref{Higgs} is devoted to the impact of the Higgs discovery on the field of particle dark matter while our summary and outlook are 
presented is section \ref{Summary}.

\section{SUSY in a nutshell}
\label{SUSYnut}
We give here the basic concepts regarding SUSY and fix the notation, which will be needed 
in the next sections. A more complete treatment of the subject can be found in \cite{Wess:1992cp,Martin:1997ns} or in the other 
reviews of this Journal. 

SUSY is a unique extension of the Poincar\'e symmetry, which is enlarged to include also
fermionic generators $Q_\alpha , Q^{\dagger\dot\alpha}$. Here we will discuss only the case
of a single additional fermionic generator and its conjugate corresponding to $N=1$ SUSY.
The new generators satisfy the algebra
\begin{equation}
\{ Q_\alpha, Q_{\dot\beta}^{\dagger} \} = - 2 (\sigma^\mu)_{\alpha \dot\beta} P_\mu
\end{equation}
and SUSY is therefore intrinsically connected to translations and 
space-time diffeomorphisms leading directly to gravity in the case of the promotion
of SUSY from a global to a local symmetry.

The principal characteristic of supersymmetric models is that every particle has to be part
of a supersymmetric multiplet consisting of different states of the same representation under
the SM gauge group, but different representation under the Lorentz group (i.e. different spin).
The basic building blocks are the chiral multiplets, here denoted by $\Phi^i$, containing
the matter fermions or scalars and their superpartners, and the vector multiplets, 
$V= g V_a T^a $ containing the gauge bosons and their superpartners, the gauginos,
one for each SM group generator $T^a$ and gauge coupling $g$.
SUSY not only offers a plethora of new particles, including DM candidates, but 
also provides a perturbative framework for extending the SM and solving some of its 
shortcomings, like the hierarchy problem, and allows for gauge coupling 
unification at a high scale and therefore points to a possible Grand Unification of the 
SM gauge group~\cite{Martin:1997ns}.

Let us turn now briefly to the case of a local supersymmetric model.
The spectrum of the theory includes not only chiral and vector multiplets, but
also a gravity multiplet, consisting of a spin-2 graviton and a spin-3/2 superpartner,
the gravitino. The Lagrangian of such a model can be written as a function of the
holomorphic superpotential $W (\Phi^i) $ and of the K\"ahler potential, $K(\Phi^i,\Phi^{*i}) $, a hermitian 
function of the chiral multiplets satisfying the gauge symmetries
of the SM model, and the gauge symmetry Killing vectors~\cite{Wess:1992cp}. 
The kinetic terms and the gauge couplings of the chiral multiplets (at lowest order in gra\-vity
and in a Minkowski background) are generated by the K\"ahler potential as
\begin{equation}
{\cal L}_{kin} =  \int d^2\theta d^2\theta^{\dagger} K_{i^{*}j} \Phi^{*i} e^{2 V} \Phi^j 
\end{equation}
where the subscript $i^{*},j$ indicate derivative with respect to the fields
$\Phi^{*i}, \Phi^j $ respectively. $K_{i^{*}j} $ is the K\"ahler metric
and gives the non-trivial structure of the scalar manifold as a 
non-linear $\sigma$-model.
The scalar potential is instead given by
\begin{equation}
{\cal V}_{pot} = e^{K/M_P^2} \left[ F_j K^{ji^{*}} F_{i^{*}} -  
3 \frac{|W|^2}{M_P^2} \right]  + \mbox{D-terms}
\label{V-scalar}
\end{equation}
where $ K^{ji^{*}} $ is the inverse metric to $ K_{i^{*}j} $ and
$ F_i = W_i + K_i \frac{W}{M_P^2} $
is the F-term corresponding to the chiral field $\Phi^i$. 
In the limit of global SUSY, i.e. $M_P \rightarrow \infty $, the potential reduces to the simple well-known expression
\begin{equation}
{\cal V}_{pot} =  \sum_i | W_i |^2  + \mbox{D-terms}
\end{equation}
assuming (as we will always do from here on) that the K\"ahler metric is canonical, 
i.e. $ K_{i^{*}j} \sim \delta_{ij} + {\cal O} (M_P^{-2}) $.

The superpotential of the MSSM contains in general quadratic and cubic interaction
terms and has the form (we adopt here the same notation of~\cite{Martin:1997ns})
\begin{equation}
W_{\rm MSSM} =  
\sbar u {\bf y_u} Q H_u -
\sbar d {\bf y_d} Q H_d -
\sbar e {\bf y_e} L H_d +
\mu H_u H_d \> 
\label{W-MSSM}
\end{equation}
where we have suppressed generation indices, but where the Yukawa couplings ${\bf y}_i$
are matrices in family space.  Here the chiral superfields $Q, L $ denote the SM $SU(2)$ doublets,
while $\sbar u$, $\sbar d$, $\sbar e$ are the $SU(2) $ singlets. $H_u$ and $H_d$ are the Higgs 
$SU(2)$ doublets chiral multiplets required to correctly achieve the SM particles mass generation and gauge anomaly cancellation.  
Note that the only dimensional term is the $\mu$-term, which has to be at the electroweak scale to allow a ``natural'' 
electroweak symmetry breaking. In a simple extension of the MSSM by a SM singlet multiplet $S$, the Next to Minimal 
Supersymmetric SM (NMSSM)~\cite{Flores:1990bt, Hugonie:2007vd, Maniatis:2009re}, such scale can be obtained 
dynamically from another cubic term and a vacuum expectation value of the singlet. The superpotential then reads
\begin{equation}
W_{\rm NMSSM} = W_{\rm MSSM}
+ \lambda S H_u H_d + \frac{1}{3} \kappa S^3 
\label{W-NMSSM}
\end{equation}
where $\lambda$ and $\kappa$ are constants. 
The gauge symmetries of the SM actually allow also additional renormali\-zable terms, which give rise to fast proton 
decay and are therefore usually forbidden by invoking an additional di\-scre\-te $Z_2$ symmetry, R-parity, which distinguishes
between particles and superparticles. Such terms are given by
\begin{equation}
W_{\rm RPB} =  \mu^{\prime i} L_i H_u + \lambda^{ ijk} L_i L_j \sbar e_k + \lambda^{\prime ijk} L_i Q_j \sbar d_k + \lambda^{\prime\prime ijk} 
\sbar u_i \sbar d_j \sbar d_k.
\end{equation}
where the couplings $\lambda^{ijk}$, $\lambda^{\prime ijk}$, $\lambda^{\prime\prime ijk}$ and $\mu^{\prime i}$ carry family indices.
If R-parity is unbroken, the lightest supersymmetric particle (LSP) is stable since R-parity forbids its decay into SM particles.
These superpotential terms must therefore vanish or be strongly suppressed to retain a supersymmetric DM candidate in the form of the LSP.
In the presence of SUSY breaking, the Lagrangian includes also soft terms, i.e. mass terms for the gauginos and the scalar fields and bilinear
and trilinear scalar terms:
\begin{eqnarray}
{\cal L}_{soft} &=& -\frac{1}{2}\sum_{i} \left( M_i \lambda_i \lambda_i  + h.c. \right)
-\stilde L^\dagger \,{\bf m^2_{L}}\,\stilde L
-\stilde {\sbar e} \,{\bf m^2_{{\sbar e}}}\, {\stilde {\sbar e}}^\dagger 
 \nonumber\\
& &
-\stilde Q^\dagger \, {\bf m^2_{Q}}\, \stilde Q
-\stilde {\sbar u} \,{\bf m^2_{{\sbar u}}}\, {\stilde {\sbar u}}^\dagger
-\stilde {\sbar d} \,{\bf m^2_{{\sbar d}}} \, {\stilde {\sbar d}}^\dagger
\nonumber\\
& &
- m^2_{H_u} H_u^* H_u - m^2_{H_d} H_d^* H_d
+ \mathcal{A} [W] 
\label{L-soft}
\end{eqnarray}
where the three sets of gauginos (one for each factor of the SM gauge group) are denoted by $\lambda_i $, with $i=1,2,3 $.  
$\stilde Q,\stilde L,\stilde{\sbar u},\stilde {\sbar d},\stilde {\sbar e}$ are the scalar superpartners in the 
multiplets $Q, L,{\sbar u},{\sbar d},{\sbar e}$ and the corresponding mass matrices are labelled with the same letters (family indices are suppressed). 
With this notation $\stilde Q_1= (\stilde u_L, \stilde d_L)$, $\stilde {\sbar u}_1 = \stilde u^{*}_R$ and similarly for the other particles~\cite{Martin:1997ns}. 
$\mathcal{A}[W]$ contains all the scalar bilinear and trilinear terms corresponding to the bilinear and trilinear terms in the superpotential $W$, e.g.
\begin{eqnarray}
\mathcal{A}[W_{\rm MSSM}] &=& 
- (\, \stilde {\sbar u} \,{\bf a_u}\, \stilde Q H_u
+ \stilde {\sbar d} \,{\bf a_d}\, \stilde Q H_d
+\stilde {\sbar e} \,{\bf a_e}\, \stilde L H_d \nonumber\\
&+& b H_u H_d + h.c. ) \,.
\end{eqnarray}
The simplest realization of the MSSM, the constrained MSSM (cMSSM), corresponds to taking
at the GUT scale a single universal mass scale $m_{1/2}$ for the gauginos, a single
mass $m_0 $ for the scalar particles and a single trilinear parameter $A$ for the three matrices ${\bf a_u},{\bf a_u}$ and ${\bf a_u}$. 
The $b$ parameter is traded for $\tan\beta = v_u/v_d$, where $v_u$ and $v_d$ are the
VEVs of the two neutral Higgs fields, while $|\mu |$ is set by the requirement of
radiative electroweak breaking.
All quantities are then extrapolated to lower energies via the relevant RGEs.
In other realizations, like the phenomenological MSSM (pMSSM), the different mass scales are disentangled and
more parameters introduced directly at the electroweak scale.

\section{SUSY dark matter candidates}
\label{candidates}

\subsection{Neutralino}
The most studied supersymmetric DM candidate is in many respects the lightest neutralino~\cite{Ellis-n0}. 
In the R-parity conserving MSSM there are four neutralinos in the mass spectrum of the theory and 
they are commonly denoted by $\tilde{\chi}^0_i$, with $i=1,\dots,4$. These mass eigenstates consist 
in four independent linear combinations involving the neutral electroweak gauginos 
($\tilde{B}$ and $\tilde{W}^{0}$) and the neutral Higgsinos ($\tilde{H}_d^{0}$ and $\tilde{H}_u^{0}$). 
The mixing between these states is a direct consequence of the electroweak symmetry breaking. 
In the gauge-eigenstate basis, represented here by the array 
$\psi=(\tilde{B},\tilde{W}^{0},\tilde{H}_d^{0},\tilde{H}_u^{0})$, the neutralino mass term in the 
MSSM Lagrangian can be writen as $-\frac{1}{2}\psi^T \mathbf{M}_{\tilde{\chi}^{0}} \psi + \textrm{h.c.}$, 
where the neutralino mass matrix $\mathbf{M}_{\tilde{\chi}^{0}}$ reads as follows~\cite{Jungman:1995df,Martin:1997ns,Bergstrom:2000pn,Bertone:2004pz}
\begin{eqnarray}
\mathbf{M}_{\tilde{\chi}^{0}} = \left(
\begin{array}{cccc}
M_1                   & 0                     & -g' v_d/\sqrt{2}  & g' v_u/\sqrt{2} \\
0                        & M_2                & g v_d/\sqrt{2}    & -g v_u/\sqrt{2}  \\
-g' v_d/\sqrt{2}   & g v_d/\sqrt{2}  &  0                      & -\mu \\
g' v_u/\sqrt{2}    & -g v_u/\sqrt{2} & -\mu                  & 0 
\end{array}
\right)
\nonumber
\,\,.
\end{eqnarray} 
The diagonal entries of this matrix, namely $M_1$ and $M_2$, stem from the gaugino mass terms present 
in the soft SUSY breaking Lagrangian (\ref{L-soft}). Because of the freedom to perform a phase 
redefinition of the fields $\tilde{B}$ and $\tilde{W}^{0}$, $M_1$ and $M_2$ can be chosen real and 
positive without loss of generality. 
Analogously, by a phase redefinition of the Higgs fields, $v_d$ and $v_u$ can be taken real and positive. 
The off-diagonal terms proportional to $\mu$ arise from the $\mu$-term in the superpotential~(\ref{W-MSSM}). 
The phase of $\mu$, which cannot be reabsorbed by further phase redefinitions, is assumed to be zero
in the vast majority of the analyses, to avoid potentially dangerous CP-violations. 
The neutralino mass matrix can be diagonalized by a unitary matrix $\mathbf{N}$ such that 
$\mathbf{N}^{*} \mathbf{M}_{\tilde{\chi}^0} \mathbf{N}^{-1} = \textrm{diag}(m_{\tilde{\chi}^{0}_1}, \dots, m_{\tilde{\chi}^{0}_4})$, where $m_{\tilde{\chi}^{0}_1}, \dots, m_{\tilde{\chi}^{0}_4}$ are the masses of the four neutralinos. The matrix $\mathbf{N}$ relates mass and gauge eigenstates as follows: $\tilde{\chi}^0_i = \mathbf{N}_{i j} \psi_j$, where in this expression the indices $i$ and $j$ label respectively mass and gauge eigenstates.
Depending on the values of the soft SUSY breaking parameters, the lightest neutralino can
also be the LSP and then a stable DM candidate for unbroken R-parity.

\begin{figure}
\vspace{1 cm}
\centering

\begin{fmffile}{fmftempl1}
  \fmfset{wiggly_len}{3mm}\fmfset{wiggly_slope}{60}
  \begin{fmfgraph*}(40,40)
   \fmfpen{0.15mm}
     \fmflabel{$\tilde{H}^{0}_u$,$\tilde{H}^{0}_d$}{f1}
     \fmflabel{$\tilde{H}^{0}_u$,$\tilde{H}^{0}_d$}{f2}	
     \fmflabel{$Z$}{l}
    \fmfleft{l}
    \fmf{photon}{l,i}
    \fmf{fermion}{f1,i,f2}
    \fmfright{f1,f2}
      \fmfset{arrow_len}{0.2 cm}\fmfset{arrow_ang}{20}
  \end{fmfgraph*}
\end{fmffile}
\hspace{1.5 cm}
~
\begin{fmffile}{fmftempl2}
  \fmfset{wiggly_len}{3mm}
  \begin{fmfgraph*}(40,40)
     \fmflabel{$\tilde{H}^{+}_u$,$\tilde{H}^{-}_d$}{f1}
     \fmflabel{$\tilde{H}^{0}_u$,$\tilde{H}^{0}_d$}{f2}	
    \fmflabel{$W^{\pm}$}{l}
    \fmfleft{l}
    \fmfpen{0.15mm}
    \fmf{photon}{l,i}
    \fmf{fermion}{f1,i,f2}
    \fmfright{f1,f2}
     \fmfset{arrow_len}{0.202 cm}\fmfset{arrow_ang}{20}
  \end{fmfgraph*}
\end{fmffile}

\vspace{1.5 cm}

\begin{fmffile}{fmftempl3bis}
\fmfcmd{
style_def ir_photon expr p =
save wiggly_len, wiggly_slope;
wiggly_len = 2 mm; wiggly_slope = 90;
draw_photon p
enddef;}
  \fmfset{wiggly_len}{3mm}
  \begin{fmfgraph*}(40,40)
    \fmflabel{$\tilde{W}^{0}$}{f1}
     \fmflabel{$\tilde{W}^{\pm}$}{f2}	
     \fmflabel{$W^{\pm}$}{l}
     \fmfpen{0.15mm}
    \fmfleft{l}
    \fmf{plain}{f1,i}
    \fmf{plain}{f2,i}
    \fmf{photon}{l,i}
    \fmf{ir_photon}{f1,i}
    \fmf{ir_photon}{i,f2}
    \fmfright{f1,f2}
     \fmfset{arrow_len}{0.2 cm}\fmfset{arrow_ang}{20}
  \end{fmfgraph*}
\end{fmffile}

\vspace{1.5 cm}

\hspace{-2.2 cm}
\begin{fmffile}{fmftempl3}
\fmfcmd{
style_def ir_photon expr p =
save wiggly_len, wiggly_slope;
wiggly_len = 2 mm; wiggly_slope = 90;
draw_photon p
enddef;}
  \fmfset{wiggly_len}{3mm}
  \begin{fmfgraph*}(40,40)
    \fmflabel{$\tilde{q}$,$\tilde{\ell}$,$H^{}_u$,$H^{}_d$,$\dots$}{f1}
     \fmflabel{$q$,$\ell$,$\tilde{H}^{}_u$,$\tilde{H}^{}_d$,$\dots$}{f2}	
     \fmflabel{$\tilde{B}$}{l}
    \fmfleft{l}
    \fmf{plain}{i,l}
    \fmfpen{0.15mm}
    \fmf{ir_photon}{l,i}
    \fmf{scalar}{f1,i}
    \fmf{fermion}{i,f2}
    \fmfright{f1,f2}
     \fmfset{arrow_len}{0.2 cm}\fmfset{arrow_ang}{20}
  \end{fmfgraph*}
\end{fmffile}
\hspace{1.5 cm}
~
\begin{fmffile}{fmftempl4}
\fmfcmd{
style_def ir_photon expr p =
save wiggly_len, wiggly_slope;
wiggly_len = 2 mm; wiggly_slope = 90;
draw_photon p
enddef;}
  \fmfset{wiggly_len}{3mm}
  \begin{fmfgraph*}(40,40)
    \fmflabel{$\tilde{q}_L$,$\tilde{\ell}_L$,$H^{}_u$,$H^{}_d$,$\dots$}{f1}
     \fmflabel{$q_L$,$\ell_L$,$\tilde{H}^{}_u$,$\tilde{H}^{}_d$,$\dots$}{f2}	
     \fmflabel{$\tilde{W}^{0}$}{l}
    \fmfleft{l}
    \fmf{plain}{i,l}
    \fmfpen{0.15mm}
    \fmf{ir_photon}{l,i}
    \fmf{scalar}{f1,i}
    \fmf{fermion}{i,f2}
    \fmfright{f1,f2}
     \fmfset{arrow_len}{0.2 cm}\fmfset{arrow_ang}{20}
  \end{fmfgraph*}
\end{fmffile}

\vspace{1.5 cm}  

\begin{fmffile}{fmftempl5}
  \fmfset{wiggly_len}{6mm}
  \begin{fmfgraph*}(40,40)
    \fmflabel{$\tilde{u}_L$,$\tilde{c}_L$,$\tilde{t}_L$}{f1}
     \fmflabel{$u^{\dagger}_R$,$c^{\dagger}_R$,$t^{\dagger}_R$}{f2}	
     \fmflabel{$\tilde{H}^{0}_u$}{l}
    \fmfleft{l}
    \fmfpen{0.15mm}
    \fmf{fermion}{l,i}
    \fmf{scalar}{f1,i}
    \fmf{fermion}{f2,i}
    \fmfright{f1,f2}
     \fmfset{arrow_len}{0.2 cm}\fmfset{arrow_ang}{20}
  \end{fmfgraph*}
\end{fmffile}
\hspace{2 cm}
~
\begin{fmffile}{fmftempl6}
  \fmfset{wiggly_len}{6mm}
  \begin{fmfgraph*}(40,40)
     \fmflabel{$\tilde{u}^*_R$,$\tilde{c}^*_R$,$\tilde{t}^*_R$}{f1}
     \fmflabel{$u_L$,$c_L$,$t_L$}{f2}	
     \fmflabel{$\tilde{H}^{0}_u$}{l}
    \fmfleft{l}
    \fmfpen{0.15mm}
    \fmf{fermion}{l,i}
    \fmf{scalar}{f1,i}
    \fmf{fermion}{f2,i}
    \fmfright{f1,f2}
      \fmfset{arrow_len}{0.2 cm}\fmfset{arrow_ang}{20}
  \end{fmfgraph*}
\end{fmffile}

\vspace{1.5 cm}

\begin{fmffile}{fmftempl7}
  \fmfset{wiggly_len}{6mm}
  \begin{fmfgraph*}(40,40)
    \fmflabel{$\tilde{d}_L$,$\tilde{s}_L$,$\tilde{b}_L$}{f1}
     \fmflabel{$d^{\dagger}_R$,$s^{\dagger}_R$,$b^{\dagger}_R$}{f2}	
     \fmflabel{$\tilde{H}^{0}_d$}{l}
    \fmfleft{l}
    \fmfpen{0.15mm}
    \fmf{fermion}{l,i}
    \fmf{scalar}{f1,i}
    \fmf{fermion}{f2,i}
    \fmfright{f1,f2}
     \fmfset{arrow_len}{0.2 cm}\fmfset{arrow_ang}{20}
  \end{fmfgraph*}
\end{fmffile}
\hspace{2 cm}
~
\begin{fmffile}{fmftempl8}
  \fmfset{wiggly_len}{6mm}
  \begin{fmfgraph*}(40,40)
     \fmflabel{$\tilde{d}^*_R$,$\tilde{s}^*_R$,$\tilde{b}^*_R$}{f1}
     \fmflabel{$d_L$,$s_L$,$b_L$}{f2}	
     \fmflabel{$\tilde{H}^{0}_d$}{l}
    \fmfleft{l}
    \fmfpen{0.15mm}
    \fmf{fermion}{l,i}
    \fmf{scalar}{f1,i}
    \fmf{fermion}{f2,i}
    \fmfright{f1,f2}
      \fmfset{arrow_len}{0.2 cm}\fmfset{arrow_ang}{20}
  \end{fmfgraph*}
\end{fmffile}
\vspace{1 cm}
\caption{Feynman diagrams for the neutralino constituents. We adopt here the same notation of Ref.~\cite{Martin:1997ns}.
}
\label{feynchi}
\end{figure}
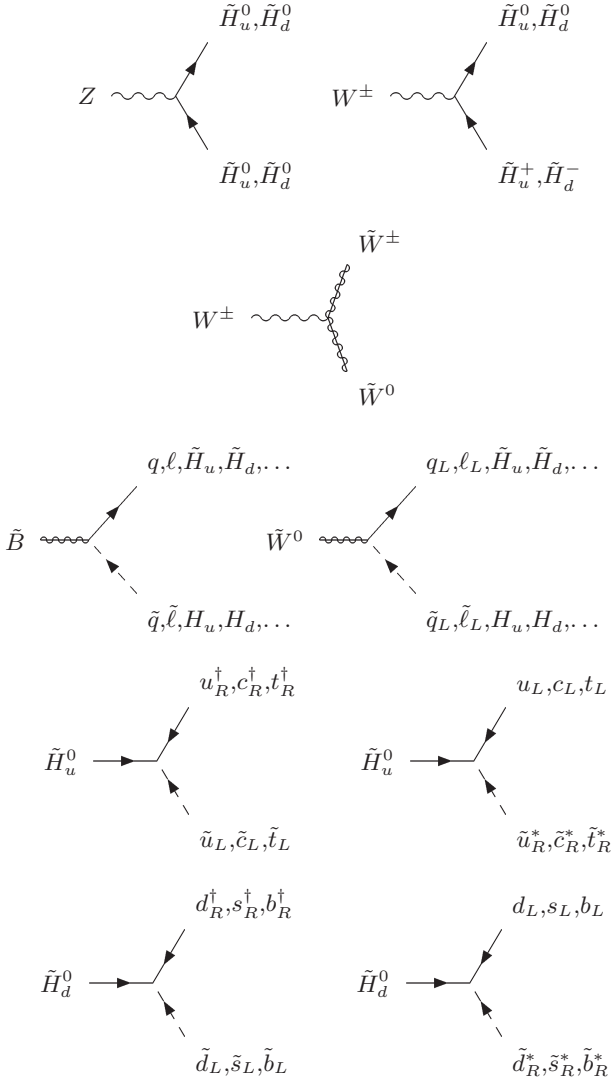

The neutralino interactions are determined by its composition ({\it i.e.} the matrix $\mathbf{N}_{i j}$), the MSSM superpotential, 
and the quantum numbers of its constituents: the $SU(2)$ singlet $\tilde{B}$, the neutral components of the $SU(2)$ doublets 
$(\tilde{H}_d^0,\tilde{H}_d^-)$ and $(\tilde{H}_u^+,\tilde{H}_u^0)$, and the neutral component 
of the $SU(2)$ triplet $(\tilde{W}^{\pm},\tilde{W}^{0})$, characterized by the hypercharges 0, -1/2, 1/2 
and 0 respectively. The gauge and Yukawa interactions allowed in the R-parity conserving MSSM for the 
neutralino constituents are shown if Fig.\ref{feynchi} in the form of Feynman diagrams. 
From these one can construct the full list of neutralino Feynman rules in the mass eigenstate basis, given for instance 
in~\cite{Dreiner:2008tw}. 

The definition of neutralino given here in the context of the MSSM can be straightforwardly generalized 
to the case of the NMSSM. In this model, the fermionic component of $S$, the ``singlino'' $\tilde{S}$, mixes with the gauge eigenstates $\tilde{H}_d^{0}$ and 
$\tilde{H}_u^{0}$. As a result, in the mass spectrum of the theory there are five neutralino-like particles 
and the lightest of them has been studied by many authors as a DM candidate (see for instance~\cite{Cerdeno:2007sn} and references therein). 
A phenomenologically interesting feature of this scenario is the existence of new interaction vertices 
(compared to the MSSM) due to the enlarged Higgs sector of the theory, which now includes 3 CP-even 
neutral Higgs bosons and 2 CP-odd neutral Higgs bosons~\cite{Maniatis:2009re}.

\subsection{Gravitino}
In local supersymmetric models we have also an electromagnetically and gauge-neutral DM candidate,
i.e. the gravitino,  the superpartner of the graviton.  
In fact, as soon as SUSY is promoted to a local symmetry, gravity is automatically included in the 
model and to complement the spin-2 graviton field, a spin-3/2 fermion must be added to the particle spectrum. 
The gravitino plays the role of  ``gauge fermion'' for SUSY and becomes massive via the SuperHiggs mechanism as 
soon as such symmetry is broken by any $F$ or $D$-term having a non-vanishing expectation value.
 The Goldstino field, providing the spin 1/2 component of the massive gravitino
 is given by a combination of the chiral fermions and gauginos along the SUSY breaking  direction singled out
by the vector $ (\langle F_i\rangle, \langle D^a \rangle) $ in field space.
The gravitino mass is in general given by
\begin{equation}
m_{3/2} = \frac{\langle |W| e^{K/(2M_P^2)} \rangle }{M_P^2}
\label{m_grav}
\end{equation}
where the brackets denote here the vacuum expectation value.
Imposing that the cosmological constant/vacuum energy in Eq.~(\ref{V-scalar}) vanishes, 
gives then, if all the D-terms vanish, also the relation
\begin{equation}
m_{3/2} = \frac{1}{\sqrt{3} M_P} \sqrt{\langle F_i F^{*}_j K^{ij^{*}} e^{K/M_P^2} \rangle}
\sim \frac{|F|}{\sqrt{3} M_P} 
\end{equation}
where $ F_i $ is the F-term of the $ith$ chiral superfield and $F$ denotes the VEV of the largest non-vanishing F-term.
In comparison, the SUSY breaking masses of the other superpartners are proportional to $F$, but can contain a
different mass scale suppression. In particular within the gauge-mediated SUSY breaking scenario~\cite{Giudice:1998bp}, the gaugino 
masses $M_i$ are given by the dominant F-term suppressed by the messenger masses, naturally smaller than 
$M_P$. In those type of models it is therefore natural to have a gravitino as the lightest supersymmetric particle. 

The gravitino couplings are dictated by gravity and SUSY and suppressed by the Planck mass as all gravity couplings.
On the other hand, the  Goldstino couples directly to the supercurrent in a derivative way and has therefore enhanced coupling in the
limit of large hierarchy between the gravitino and the other sparticle masses. The general gravitino couplings can be found 
in~\cite{Wess:1992cp, Pradler:2007ne,Grefe:2011dp}.
\subsection{Axino}
Another neutral superparticle that can play the role of DM is the axino, the superpartner of the axion field. It arises naturally
in extensions of the SM including also the Peccei-Quinn~\cite{Peccei:1977hh} solution to the strong CP problem in a
supersymmetric setting~\cite{Nilles:1981py,Frere:1982sg}. The axino is a spin-1/2 fermion and it is univocally defined (and nearly 
massless) only in the limit of unbroken SUSY~\cite{Covi:2009pq}. 
In that case in fact the whole axion supermultiplet, including the axino and the saxion as scalar partner of the pseudoscalar axion, 
is protected by the Goldstone nature of the axion and it is massless as long as one neglects the explicit symmetry breaking coming 
from QCD instantons effects. On the other hand, as soon as SUSY is broken, the axino acquires a mass and also mixes with the other
neutral fermions rendering its mass and phenomenology strongly model dependent. Note that some axion models 
of the DFSZ-type \cite{Dine:1981rt,Zhitnitsky:1980tq} introduce an axion coupling to the Higgs fields in a similar way to the singlino 
couplings in the NMSSM, mixing in general the axino with the neutralinos,  but the two models differ in the presence of cubic or 
quadratic couplings for the singlet field. 

If the main axion/axino couplings are only with the QCD sector, as it happens instead in the KSVZ-type models~\cite{Kim:1979if,Shifman:1979if}, 
the neutralino mass matrix retains an eigenstate strongly aligned with the axion direction and decoupled from the rest of the spectrum. 
In that case the phenomenology simplifies as one can approximate the axino couplings with the supersymmetrized axion ones~\cite{Covi:2001nw}.
Note that this requires to extrapolate the axion couplings to high scale, which may not always be possible \cite{Bae:2011jb, Choi:2011yf}.
Then the axino couplings are suppressed by the axion decay coupling  $ f_a$, which is constrained by axion physics 
considerations~\cite{Raffelt:2006cw} to lie between $ 10^9-10^{12} $ GeV. The axino is therefore naturally very weakly interacting and
can be a realization of the SuperWIMP mechanism if the reheat temperature is very low \cite{Covi:1999ty,Covi:2001nw} 
or be produced by thermal decays or scatterings~\cite{Covi:2001nw,Covi:2002vw,Brandenburg:2004du,Strumia:2010aa,Chun:2011zd,Bae:2011jb,Choi:2011yf}.
\subsection{Other candidates}
\label{others}
A variety of DM models have been developed in the lite\-ra\-ture where the DM properties are to some extent influenced by the ones of the SM neutrinos. 
The MSSM left-handed sneutrino $\tilde{\nu}$ has been excluded long ago as DM candidate because 
of its sizable coupling with the $Z$ boson, which leads to a too large annihilation cross section 
(implying a too small relic density) and a too large DM-nucleon scattering cross section, 
which is experimentally excluded~\cite{Falk:1994es}. 
One possibility to reconcile $\tilde{\nu}$ DM with observations is to
add to the MSSM spectrum a new superfield 
(for each neutrino flavor) whose bosonic component, the right-handed sneutrino, mixes with the 
left-handed sneutrino~\cite{Hall:1997ah}. 
This mixing reduces the strength of the dangerous $\tilde{\nu}$-$\tilde{\nu}$-$Z$ coupling, eventually leading to an acceptable phenomenology for the 
lightest sneutrino mass eigenstate, which in this context qualifies as a viable scalar DM candidate. 
A Higgs boson mass of 125-126 GeV restricts the allowed supersymmetric configurations to regions in parameter space where the mixed sneutrino has 
a mass of the order of 100 GeV~\cite{Dumont:2012ee}. At the same time in this class of theories different implementations of the seesaw mechanism provide 
a procedure to generate the correct masses for the light SM neutrinos (see for instance~\cite{Arina:2008bb}). On the other hand the coupling to the 
$Z$ boson of a pure right-handed sneutrino is exactly zero, a property which makes the right-handed sneutrino phenomenologically safe as 
non-thermal ``FIMP''-like DM candidate~\cite{Asaka:2005cn}, 
but at the same time essentially impossible to detect if the underlying theory is the MSSM. A phenomenologically more 
interesting DM candidate is the right-handed sneutrino in the context of the NMSSM~\cite{Cerdeno:2008ep}. In this 
case the coupling of the right-handed sneutrino to the Higgs bosons is enhanced by new interactions which are controlled by the extra 
parameters of the superpotential (\ref{W-NMSSM}) and therefore make this DM candidate potentially detectable by the next generation of 
ton-scale direct detection experiments. 

We conclude this section with an alternative to the WIMP paradigm: the WIMPless scenario~\cite{Feng:2008ya}. 
In this class of SUSY theories the field content is divided in three sectors: a visible (or MSSM) sector, the SUSY breaking sector 
and, finally, the hidden sector which contains the DM candidate. In these models 
the SUSY breaking is mediated to the hidden sector by gauge interactions of arbitrary strength $g_\chi$. 
As for the familiar gauge-mediated SUSY breaking mechanism, 
this setup implies that the DM candidate acquires after SUSY breaking a mass of the order of 
$m_\chi \sim g_\chi^2 F/M$, where the parameters $F$ and $M$ parameterize the F-term and VEV of the SUSY 
breaking field. Assuming a standard thermal history in the hidden sector, the present DM density for this model scales as
$\Omega_\chi \sim (H^{*}/n^{*}_{\chi})(m_\chi^2/g_\chi^4)$, where $H$ and $n_{\chi}$ are evaluated at the time $t^{*}$ of the DM chemical decoupling. 
Hence in this framework $\Omega_\chi$ depends on $F$ and $M$ only, for what concerns particle 
physics. This implies that in this class of theories the relic density matches the observations for a broad range of DM couplings and masses, 
making this scenario phenomenologically very interesting.

\section{SUSY dark matter detection}
\label{detection}
In the following we will review the current status of SUSY DM searches focusing separately on two 
different classes of DM candidates. First we will concentrate on the neutralino which is the
archetypal WIMP and has couplings of 
electroweak strength and a mass typically varying between 
a few GeV up to tens of TeV. 
Then we will move to a second class of particles, like the gravitino, 
whose interactions are much weaker than the electroweak force 
({\it e.g}. in principle down to gravitational strength) and that are 
characterized by a larger range of possible masses. These candidates are 
commonly dubbed SuperWIMPs.  

\subsection{SUSY WIMPs}
\subsubsection{Direct Detection}
\begin{figure}
\resizebox{0.48\textwidth}{!}{%
  \includegraphics{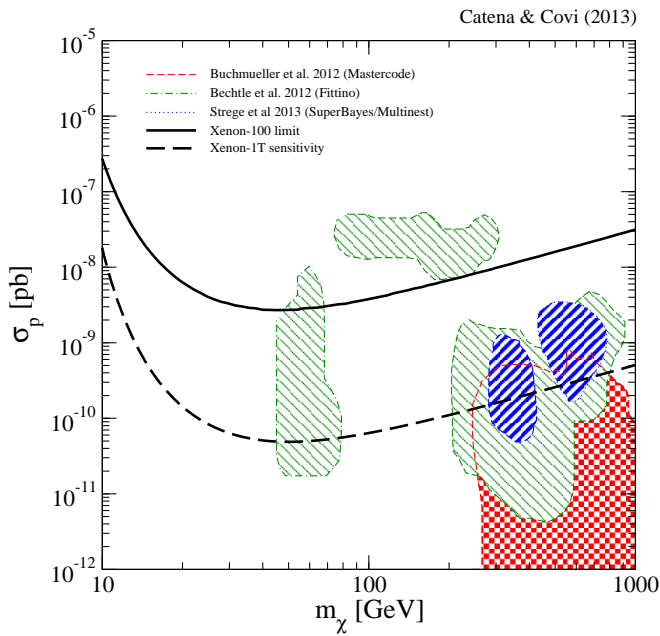}
}
\caption{Regions in the plane DM mass $m_{\chi}$ vs spin-independent DM-nucleon scattering cross section $\sigma_p$ favored by three independent global fits of the MSSM to a variegated sample of observations including the latest LHC data, low energy observables, cosmological limits as well as DM searches. The blue contours include the 95\% credible regions found in Ref.~\cite{Strege:2012bt} employing log-priors, while the red and green countours represent the 95\% C.L. favored regions determined in 
Refs.~\cite{Bechtle:2012zk,Buchmueller:2012hv}. The solid black line corresponds to the present XENON100 exclusion limit. The black dashed curve represents instead the sensitivity of XENON-1T after one year of data.}
\label{DD}       
\end{figure}
The aim of the direct detection technique~\cite{Goodman:1984dc} is to measure the energy released by Milky Way DM particles when scattering off nuclei in underground detectors. In the case of DM-nucleus spin-independent interactions, the differential rate of scattering events expected in a direct detection experiment, $dR/dQ$, is given by~\cite{Jungman:1995df,Bergstrom:2000pn,Bertone:2004pz}:
\begin{equation}
\frac{dR}{dQ}\left(Q\right) =\gamma(Q)\frac{\sigma_p}{m_\chi}\int_{|\vec{v}|<v_{min}} d^3 v \, \frac{f_{\odot}(\vec{v},t)}{|\vec{v}|}
\label{rate}
\end{equation}
where $f_{\odot}$ is the time dependent local DM distribution function in the detector rest frame normalized at the value of the local DM density, 
$v_{min}(Q,m_\chi)$ is the minimum DM velocity required to transfer an energy $Q$ to the detector nuclei, $m_\chi$ is the DM mass and 
$\sigma_p$ the DM-nucleon scattering cross section. The energy dependent function $\gamma(Q)$ incorporates the details of the 
detector. 

The calculation of the DM-nucleus scattering cross section has been performed in various supersymmetric extensions of the SM (and we refer the reader to 
Ref.~\cite{Drees:1993bu,Ellis:2000ds} for details regarding these computations). In the vast majority of the proposed scenarios - for instance in the case of a 
neutralino within the MSSM - this calculation reduces to the determination of the couplings, often denoted by $\alpha_{2q}$ and $\alpha_{3q}$, appearing in the contact interactions 
\begin{equation}
\alpha_{3q}\,\bar{\chi}\chi\bar{q}q \qquad \textrm{and} \qquad \alpha_{2q}\,\bar{\chi}\gamma^{5}\gamma^{\mu}\chi\bar{q}\gamma^{5}\gamma_{\mu}q\,
\label{eq:ddoperators}
\end{equation}
which in the non-relativistic limit lead to a spin-in\-de\-pen\-dent and to a spin-dependent DM-nucleus interaction respectively. In the MSSM, these operators are the only velocity independent operators relevant for the neutralino-nucleus scattering. For Dirac DM candidates, instead, a vector coupling of the form $\bar{\chi}\gamma^{\mu}\chi\bar{q}\gamma_{\mu}q$ is also allowed. A complete classification of the non relativistic operators allowed by Galilean invariance as well as by energy and momentum conservation and relevant for the DM-nucleus scattering can be found in~\cite{Fitzpatrick:2012ib}. 

Several experiments have currently reached the sensitivity to start probing the WIMP paradigm using different target materials and detection principles, including the detection of an annual modulation signal associated with the time variation of the expected DM scattering rate in the detector due to the motion of the Earth relative to the Sun. Three collaborations, namely CRESST, CoGeNT and CDMS have also published results compatible with the detection of a small number of candidate signal events, which were not possible to ascribe to any of the considered background sources~\cite{Angloher:2011uu,Aalseth:2011wp,Agnese:2013rvf}. There is not a general consensus regarding the interpretation of these results and the picture is further complicated by the 8.9~$\sigma$ C.L. detection of a modulation signal made by the DAMA/LIBRA experiment~\cite{Bernabei:2010mq}. This finding has been not confirmed by other experiments and its interpretation in terms of dark mass and scattering cross section is in very strong tension with the results of other experiments, in particular of XENON100~\cite{Aprile:2012nq}, which is currently excluding the regions of the plane DM mass versus spin-independent scattering cross section favored by DAMA as well as the low WIMP mass regions favored by CRESST, CoGeNT, and CDMS. A neutralino with a mass close to 10 GeV, as required by these experiments, might be generated by relaxing the assumption of gaugino mass unification~\cite{Fornengo:2010mk} while DM isospin violating interactions seem the only possibility to reconcile DAMA with XENON100~\cite{Feng:2011vu}. In addition, several experiments are also probing the spin-dependent DM interactions using nuclei with unpaired protons as target materials ({\it e.g.} see Ref.~\cite{Kim:2012rza}). 

The impact of these results on the search for SUSY WIMPs is remarkable. In Fig.~\ref{DD} we report the regions in the plane DM mass $m_{\chi}$ versus spin-independent DM-nucleon scattering cross section $\sigma_p$ favored by three independent analyses of the MSSM~\cite{Strege:2012bt,Bechtle:2012zk,Buchmueller:2012hv}. 
Several supersymmetric configurations appear already excluded by current direct detection searches. Moreover, the next generation of ton-scale experiments will be able to probe the vast majority of the presently allowed configurations. However, DM candidates with a mass in the 10 GeV (100 GeV) range and spin-independent cross sections smaller than roughly $10^{-45}$ cm$^2$ ($10^{-49}$ cm$^2$) will be very difficult (if not impossible) to discover even with the next generation of direct detection experiments, since an experiment sensitive to such a low scattering cross sections would also measure the large flux of solar and atmospheric neutrinos which would therefore constitute a copious and irreducible background source~\cite{Billard:2013qya}.

The interpretation of a DM direct detection signal in terms of DM properties is significantly affected by the uncertainties in local DM density~\cite{Catena:2009mf}, in the local DM velocity distribution function~\cite{Catena:2011kv,Bozorgnia:2013pua} and by the current limited knowledge of the nuclear form factors as well as of the matrix elements determining the DM-nucleon couplings~\cite{Cerdeno:2012ix}. For these reasons DM halo independent approaches to the WIMP direct detection have been proposed~\cite{Fox:2010bu,Fox:2010bz,DelNobile:2013cva,HerreroGarcia:2012fu} as well as multiple targets analyses~\cite{Pato:2010zk}. 

\subsubsection{Indirect Detection}
\label{id}
Alternatively, DM could be revealed through the observation of SM particles produced in 
space by DM annihilations or decays~\cite{Bergstrom:2000pn,Bertone:2004pz} (the latter possibility applies to long-lived DM candidates and will be reviewed in section~\ref{superwimps}). This detection 
strategy is known as the DM indirect detection technique. WIMPs are 
expected to copiously annihilate in galactic and extragalactic regions where 
the DM density is large compared to the present mean cosmic density. 
Annihilation products of particular interest are $\gamma$-rays~\cite{Bringmann:2012ez} - which 
propagate along geodesics and provide therefore direct information on the 
region where the associated annihilations have occurred - and antimatter~\cite{Salati:2010rc}
({\it e.g.} positrons, antiprotons or exotic nuclei like antideuterons~\cite{Donato:1999gy}, etc\dots), which is also 
produced by standard astrophysical sources ({\it e.g.} pulsars) but 
nevertheless sub-leading in space. In both cases the expected energy spectra 
exhibit a kinematical cutoff associated with the mass of the DM 
candidate, a feature which can be used to separate the DM signal 
from possible astrophysical backgrounds which (with a few exceptions)
are characterized by power laws decreasing with energy. 

The flux of charged annihilation products observable on top of the atmosphere is calculated solving a transport equation for the propagation and diffusion of these particles in the galactic magnetic field. This can be done either numerically~\cite{Strong:1998pw} or semi-analitically, expanding in Bessel functions the space and energy distribution function of the DM annihilation products~(see for instance Ref.~\cite{Delahaye:2007fr}). For neutral annihilation products of type $i$, such as photons and neutrinos, the observable differential flux in a direction at an angle $\theta$ from the direction of the galactic center, at an energy $E$ is given by~\cite{Bergstrom:2000pn,Bertone:2004pz}:
\begin{equation}
\frac{d\Phi_i}{dE  d\Omega}(\theta, E) = \frac{1}{4\pi m_{\chi}^2} \langle \sigma v \rangle \frac{dN_i}{dE} (E)
\int_{l.o.s.} ds \rho_{\chi}^2(r(s,\theta))
\label{eq:Phi}
\end{equation}
where $\langle \sigma v \rangle $ is the average velocity-weighted annihilation cross section, $\frac{dN_i}{dE}$ is the differential energy spectrum of the $i$ particles produced per annihilation and the integral of the squared DM mass density, $\rho_{\chi}^2$(r), is computed along the line of sight $s$, where $r(s,\theta)$ is the distance from the galactic center. Therefore, we clearly see that regions in space with a high concentration of DM are good targets to look for such annihilations.
For Majorana DM candidates Eq.~(\ref{eq:Phi}) has to be divided by a factor of 2. 

$\gamma$-rays from DM annihilations can be produced either through a prompt emission or as a byproduct of various processes, including the hadronization of charged annihilation products forming a $\pi^{0}$ subsequently decaying in a pair of photons and the inverse Compton of relativistic charged particles upscattering low energy photons from CMB, starlight and interstellar radiation~\cite{Cirelli:2009vg}.
 
A study of the signal to noise ratio shows that optimal targets are~\cite{Kuhlen:2012ft}: the galactic center (with a large expected DM signal 
but also a large astrophysical background), dwarf spheroidal galaxies (which are among the most DM dominated environments), the galactic halo (including the associated substructures) and, finally, massive nearby galaxy clusters. DM can be also gravitationally trapped in astrophysical objects like the Sun and annihilate at their center producing a potentially observable flux of energetic neutrinos~\cite{Bergstrom:1998xh}.
\begin{table}
\caption{s-channel and t-channel tree level two-body neutralino annihilations allowed in the MSSM~\cite{Bertone:2004pz}.
}
\label{ann}       
\begin{tabular}{lllllll}
\hline\noalign{\smallskip}
final state & &  & &  s-channel &&  t-channel \\
\noalign{\smallskip}\hline\noalign{\smallskip}
$\tilde{\chi}^{0}_1\tilde{\chi}^{0}_1 \rightarrow \bar{f} f$ & & & &  $A$,$Z$ & & $\tilde{f}$ \\
$\tilde{\chi}^{0}_1\tilde{\chi}^{0}_1  \rightarrow W^{\pm}W^{\mp}$  & & & & $h$,$H$,$Z$  & & $\tilde{\chi}^{\pm}_i$\\
$\tilde{\chi}^{0}_1\tilde{\chi}^{0}_1  \rightarrow Z Z $ & &  & &  $h$,$H$ & & $\tilde{\chi}^{0}_i$\\
$\tilde{\chi}^{0}_1\tilde{\chi}^{0}_1  \rightarrow Z A $ & &  &   & $h$,$H$,$Z$ & &$\tilde{\chi}^{0}_i$\\
$\tilde{\chi}^{0}_1\tilde{\chi}^{0}_1  \rightarrow Z h (Z H)$ &  & & &  $A$,$Z$ & & $\tilde{\chi}^{0}_i$\\
$\tilde{\chi}^{0}_1\tilde{\chi}^{0}_1  \rightarrow W^{\pm} H^{\mp} $ & & & &  $A$,$h$,$H$ & & $\tilde{\chi}^{\pm}_i$\\
$\tilde{\chi}^{0}_1\tilde{\chi}^{0}_1  \rightarrow hh (HH) $ & &  & & $h$,$H$ & & $\tilde{\chi}^{0}_i$\\
$\tilde{\chi}^{0}_1\tilde{\chi}^{0}_1  \rightarrow AA $ & &  & & $h$,$H$ & & $\tilde{\chi}^{0}_i$\\
$\tilde{\chi}^{0}_1\tilde{\chi}^{0}_1  \rightarrow Ah (AH) $ & & & &  $A$,$Z$ & & $\tilde{\chi}^{0}_i$\\
\noalign{\smallskip}\hline
\end{tabular}
\end{table}

The DM annihilation cross section has been calculated in various supersymmetric models. In the MSSM there are different classes of two-body final states allowed at tree level~\cite{Drees:1992am,Bergstrom:2000pn,Bertone:2004pz} (we neglect here coannihilations, which are instead crucial to calculate the DM relic density). A complete list is shown in Tab.~\ref{ann}, where we assumed that the neutralino is the DM candidate. In addition, there are also three body final states which can play an important role in the context of the DM indirect detection, in particular those involving a single photon or electroweak gauge boson emitted as final state radiation or by the particle mediating the annihilation (the so-called virtual internal bremstrahlung) which can lead to interesting spectral features in the $\gamma$-ray band~\cite{Bringmann:2012vr}. Detailed calculations of the associated annihilation cross sections have been presented by various groups~\cite{Bringmann:2007nk,Ciafaloni:2011sa} and also implemented in publicly available numerical codes~\cite{Gondolo:2004sc}. At one loop a pair of photons~\cite{Bergstrom:1997fh} or one photon accompanied by one $Z$-boson~\cite{Ullio:1997ke} can be also produced. These monochromatic lines are clearly distinguishable by standard astrophysical backgrounds but unfortunately loop-suppressed.

The last four years of observations in the field of indirect DM searches have been particularly rich of interesting results. Regarding DM searches in the antimatter channel, the observation of an ``anomalous'' rise in the positron fraction measured by the PAMELA satellite~\cite{Adriani:2008zr} and possibly related to a nearby primary source of positrons has triggered a vigorous debate in the literature concerning the DM interpretation of this signal. Though pulsars might explain this observation as well~\cite{Hooper:2008kg,Linden:2013mqa}, a clear and universally accepted interpretation of this phenomenon is still missing. Recently, a rise in the positron fraction in the 10 GeV - 300 GeV range has been also observed by the AMS-02 space observatory~\cite{Aguilar:2013qda}, providing therefore a remarkable confirmation of the PAMELA results. Likely, future data from the AMS-02 experiment will finally clarify this intriguing puzzle. The PAMELA satellite has also provided accurate measurements of the antiproton flux on top of the atmosphere~\cite{Adriani:2010rc} which are however in excellent agreement with expectations from standard astrophysical sources. In addition, during the past few years $\gamma$-ray observations have also played a major role in the context of DM searches. The identification of a 130 GeV $\gamma$-ray line in the direction of the galactic center in the Fermi-LAT data~\cite{Bringmann:2012vr,Weniger:2012tx} has stimulated an intense discussion regarding its nature. Though a monochromatic $\gamma$-ray line at these energies has been often referred to as the ``smoking gun'' for DM searches, the global significance of this signal is constantly decreasing (currently quoted at 1.5 $\sigma$~\cite{Fermi-LAT:2013uma}) while the amount of data taken is increasing. Significant results have been also achieved observing the galactic center with HESS, a system of imaging atmospheric Cherenkov telescopes designed to investigate cosmic $\gamma$-rays in the energy range from 10's of GeV to 10's of TeV. Finally, observations in the radio band have also provided valuable results for DM searches through the synchrotron emission generated by relativistic charged particles produced by DM matter annihilations in the galactic magnetic field~\cite{Fornengo:2011cn}. 

These data significantly limit the allowed regions in the plane DM mass versus annihilation cross section and consequently drastically impact the space of allowed supersymmetric configurations. For instance, the latest Fermi-LAT data from the observation of 25 dwarf spheroidal satellite galaxies constrain the DM annihilation cross section to be less than $3\times10^{-26}$~cm$^3$~s$^{-1}$ (``the thermal cross section'') for DM particles with a mass less than 10 GeV (15 GeV), assuming that the dominant annihilation channel is $\bar{b}b$ ($\tau^+\tau^-$)~\cite{Ackermann:2013yva}. The thermal cross section is commonly taken as an important reference when extracting bounds from indirect detection measurements, since it represents the cross section required for thermal WIMPs to match the observed relic density. This value can be however drastically different if the expansion rate of the Universe before DM decoupling is not the one expected from General Relativity, as typically occurs for instance in Scalar-Tensor theories of gravity and in certain models with extra spatial dimensions~\cite{Catena:2004ba}. Indeed, a larger expansion rate, for example, implies an earlier DM chemical decoupling and therefore a larger DM density at decoupling which has thus to be compensated by a larger annihilation cross section~\cite{Catena:2009tm,Schelke:2006eg}. This phenomenon induces a distortion in the allowed regions of the ($m_{1/2}$,$m_{0}$) plane of the cMSSM~\cite{Catena:2007ix}.
\subsubsection{LHC searches}
\label{wimpLHC}
WIMPs could be also produced in high energy proton-proton collisions at the LHC. Different final states are in principle relevant for DM searches.
One of the most popular channel to look for DM is a final state involving a single jet or a single photon produced in association with missing transverse momentum. This class of processes is generically expected in all models where there is an effective contact interaction involving two DM particles and two quarks (not only in SUSY theories). The single jet (or the single photon) is radiated by one of the initial state quarks. In the case of DM searches in mono-jet (and analogously for the mono-photon) events, the most relevant source of SM background consists in the production of a $Z$ boson in association with a jet, with the $Z$ boson decaying into a neutrino pair, or in the $W$ plus jet production, with the $W$ boson decaying into a neutrino and a misidentified lepton. Current measurements performed by the ATLAS and CMS collaborations focused on these channels are consistent with SM expectations~\cite{CMS-PAS-EXO-12-048,Aad:2012fw}. A similar strategy currently pursued at LHC consists in searching for DM in events with a hadronically decaying $W$ or $Z$ boson. Also the study of this channel has however reported results which are consistent with SM expectations~\cite{Aad:2013oja}. More recently, it has been proposed to search for a DM signal in mono-lepton events resulting from the production of a pair of DM particles in association with a $W$ boson subsequently decaying into a lepton and a neutrino. The latest analysis of this channel performed using data corresponding to 20 fb$^{-1}$ of integrated luminosity at $\sqrt{s}=8$ TeV center-of-mass-energy has found no indications of a DM signal~\cite{CMS-PAS-EXO-13-004}.

In addition to these search strategies, which as already mentioned would also apply to non SUSY WIMPs, there are different studies which instead aim at identifying DM explicitly assuming that the underlying theory is SUSY. An interesting example of this type of analyses is the search for neutralino DM in events involving direct slepton or gaugino production in final states with two or three leptons and missing transverse momentum~\cite{ATLAS-CONF-2013-049,ATLAS-CONF-2013-035}. In the case of the direct production of sleptons, which occurs via a supersymmetric version of the Drell-Yan process, the following chain of decays leads to neutralino DM production: $\bar{q}q \rightarrow \tilde{\ell}^{\pm}\tilde{\ell}^{\mp} \rightarrow (\tilde{\chi}^{0}_1\ell^{\pm}) +(\tilde{\chi}^{0}_1\ell^{\mp})$. In the case of direct gaugino production, instead, DM can be produced in various way. As an example, we mention here the decay chain:  $\bar{q}q \rightarrow \tilde{\chi}^{0}_2 \tilde{\chi}^{\pm}_1 \rightarrow (\tilde{\chi}^{0}_1\ell^{\pm}\ell^{\mp}) +  (\tilde{\chi}^{0}_1\ell^{\pm}\nu)$, where DM is produced in association with three leptons in the final state. Again, current searches are consistent with SM expectations~\cite{ATLAS-CONF-2013-049,ATLAS-CONF-2013-035}. The same conclusion also applies to another channel relevant for SUSY DM, namely the search for strongly produced supersymmetric particles in decays with two leptons and missing transverse momentum~\cite{ATLAS-CONF-2013-089}.

There are two main approaches to extract from these LHC searches limits on the DM mass and couplings. A first possibility consists in modeling the DM interactions within an effective field theory framework and then assuming that only certain operators are relevant when studying DM at collider~\cite{Goodman:2010ku}. This approach has the advantage of establishing explicit correlations between LHC observables, {\it e.g.} missing transverse momentum distributions, and other DM properties, such as for instance the DM-nucleon scattering cross section. The drawback of this approach is that it might provide an oversimplified picture of the real DM properties~\cite{Buchmueller:2013dya}. Alternatively the LHC searches for DM can be interpreted within specific particle physics framework, like for instance the pMSSM, where the SUSY spectrum is described by approximately $\mathcal{O}(10)$ parameters. In this case the large number of parameters tends to weaken the possibility of directly relating the LHC results to other DM detection strategies. Nevertheless, focusing on certain classes of supersymmetric configurations, interesting correlation patterns have been identified even in the pMSSM framework~\cite{Arcadi:2012uh}.
\subsubsection{Complementarity of the different detection strategies}
The detection strategies presented in this section probe distinct WIMP properties and are therefore complementary. This allows on the one hand to independently verify a hypothetical DM discovery made by one of the mentioned experimental techniques, on the other hand to experimentally probe a large set of different DM models.

Direct detection experiments and LHC searches can be combined in different ways. One can combine the associated measurements in a global fit to infer the properties of the underlying DM model \cite{Strege:2012bt} or, alternatively, use the results from one of these detection strategies to predict, within a certain particle physics framework, DM signals in the other class of experiments. For instance, LHC (real and simulated) data have been often used in the literature to try to reconstruct the DM-nucleon scattering cross section within the MSSM~\cite{Baltz:2006fm}. There have been recently also attempts to use simulated results from the next generation of ton-scale direct detection experiments to forecast certain classes of missing energy distributions observable at the LHC~\cite{Arcadi:2012uh}. In this case, the basic idea is that although the direct detection technique is directly sensitive only to the DM mass and scattering cross section, indirectly this class of experiments has also the potential to constrain the parameters, or certain combinations of parameters, which most crucially enter the calculation of specific DM production cross sections at the LHC. An illustrative example of this approach can be found in Ref.~\cite{Arcadi:2012uh}, where this idea is applied to a light neutralino thermally produced in the early universe via resonant annihilations mediated by the CP-even Higgs boson. In this scenario a hypothetically discovered direct detection signal can be translated into a prediction for 
the missing energy distribution associated with a LHC final state involving three leptons and missing energy. Similar approaches to the ones presented here have been also investigated to perform combined analyses of DM searches at LHC and in space via $\gamma$-ray observations~\cite{Bertone:2011pq,Kopp:2013mi}.

Direct and indirect detection searches are also highly complementary~\cite{Bergstrom:2010gh,Arina:2013jya}. In the MSSM this has been clearly shown 
in Ref.~\cite{Bergstrom:2010gh}, where through a scan of the cMSSM parameter space the authors prove that in the plane $\langle \sigma v \rangle/m_\chi^2$ versus spin-independent scattering cross section $\sigma_{p}$, direct and indirect detection experiments probe orthogonal regions. Focusing for instance on the expected sensitivity of the next-generation of imaging Cherenkov telescope arrays and assuming a set of dwarf spheroidal galaxies as target regions, in Ref.~\cite{Bergstrom:2010gh} it is shown that portions of the cMSSM parameter space corresponding to values of $\langle \sigma v \rangle/m_\chi^2$ smaller than $10^{-31}~\textrm{cm}^3~\textrm{s}^{-1}\textrm{GeV}^{-2}$ (and to $\sigma_p>10^{-11}~\textrm{pb}$ ) will be explored by ton-scale direct detection experiments but not by indirect searches, on the contrary, regions with $\sigma_{p}<10^{-11}~\textrm{pb}$ will be probed by indirect searches (if $\langle \sigma v \rangle/m_\chi^2>10^{-31}~\textrm{cm}^3~\textrm{s}^{-1}\textrm{GeV}^{-2}$) remaining inaccessible to direct detection experiments. 

\subsection{SUSY SuperWIMPs}
\label{superwimps}
In the case of particles with interactions much weaker than the electroweak one, the chance of DM detections are much more limited
than for WIMPs. Nevertheless in particular models where the DM candidate is unstable or couples with more strongly interacting
particles, like in the case of gravitino or the axino, the situation is still promising.
\subsubsection{Direct detection}
The direct detection of particles like the gravitino is unfortunately very difficult, since the scattering cross section is strongly suppressed
and practically always below the unavoidable neutrino background. In fact the elastic scattering of a gravitino against the nucleus
must proceed through the supergravity dimension-six four-fermion contact interaction or through two single-gravitino vertices,
giving a rate suppressed by four powers of the Planck mass.
In models where R-parity is violated, also the inelastic scattering of the gravitino into a neutrino is possible, which is instead
suppressed only by two powers of the Planck scale and the smallness of the R-parity coupling. Unfortunately for values of such coupling
compatible with indirect detection bounds, the rate is also in this case unobservable \cite{Grefe:2011dp}.
Also the axino interaction with quarks is unfortunately too strongly suppressed to give rise to a measurable signal.
We can conclude therefore that a confirmed DM signal in a direct detection experiment would be very difficult to reconcile with
the hypothesis that DM is a gravitino or axino.
\subsubsection{Indirect detection}
It has been realized few years ago that gravitino LSP and other SuperWIMPs can be retained as good DM candidates
even if R-parity is slightly broken and the LSP is no more stable~\cite{Takayama:2000uz, Buchmuller:2007ui}. 
In fact in most cases the lifetime of the DM particle can still be long enough to exceed the lifetime
of the Universe by many orders of magnitude. In the case of the gravitino such small R-parity breaking would
be actually quite welcome, since it allows to avoid any Big Bang Nucleosynthesis (BBN) constraint coming from the 
(too) late decay of the Next to Lightest Supersymmetric Particle (NLSP)~\cite{Jedamzik:2009uy}.
Indeed in the presence even of a tiny violation coupling above $10^{-12}-10^{-10}$, most of the NLSPs decay quickly in SM
particles before BBN~\cite{Buchmuller:2007ui}. For other DM candidates like the axino the BBN constraints are
less severe~\cite{Baer:2010kw, Baer:2010gr, Freitas:2011fx} and therefore the introduction of R-parity breaking perhaps less compelling, 
but still even in that case the possibility of a decaying DM candidate remains open \cite{Covi:2009pq}.
Note that, on the other side, the R-parity breaking couplings cannot be too large if one wants to retain
the baryonic asymmetry~\cite{Dreiner:1992vm,Endo:2009cv}.

If the DM is not stable, we may be able to detect its decay in our galactic neighborhood or from any of the
astrophysical targets discussed already in section~\ref{id}. The differential flux of SM particles produced by decaying DM is given by:
\begin{equation}
\frac{d\Phi_i}{dE d\Omega}(\theta, E) = \frac{1}{4\pi\tau_{\chi}m_{\chi}} \frac{dN_i}{dE} (E)
\int_{l.o.s.} ds \rho_{\chi}(r(s,\theta))
\end{equation}
where $\tau_{\chi}$ is the DM decay time, $\frac{dN_i}{dE}$ is the differential energy spectrum of the SM particles 
of type $i$ and energy $E$ produced per decay and the last integral is computed along the line of sight.
From this expression, we expect from decaying DM a very different spatial distribution of the signal compared to the 
annihilation case. Moreover the strength of the signal is not very strongly dependent on the particular DM profile density and 
therefore the bounds are much less affected by astrophysical uncertainties~\cite{Bertone:2007aw}.
Note also that, contrary to the case of WIMPs, there is here no direct connection between the DM decay and the DM production 
mechanism and therefore there is no natural expectation for the DM lifetime.

\begin{figure}
\resizebox{0.48\textwidth}{!}{%
  \includegraphics{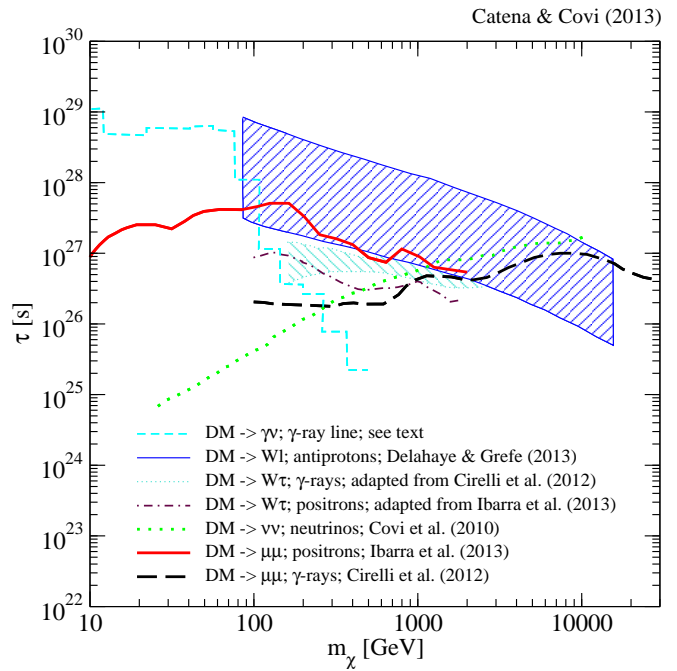}
}
\caption{A compilation of 2-$\sigma$ exclusion limits in the plane DM mass versus mean DM lifetime for different decay channels.
For the gravitino decay in models with bilinear R-parity violation, 
we give the limits from Fermi-LAT $\gamma$-ray line searches from \cite{Vertongen:2011mu} 
rescaled according to the gravitino branching ratio in \cite{Covi:2008jy,Grefe:2011kh} 
as the light blue (short-dashed) line. Moreover for the same gravitino decaying DM model the blue band
corresponds to the limits from antiprotons recently computed in \cite{Delahaye:2013yqa}.
The size of this band incorporates the uncertainties in the assumed galactic diffusion model. 
As an indication of the reach of other channels for the same gravitino models, we also give a 
conservative estimate on the bounds on the dominant gravitino decay channel $W\tau$ combining the $WW$ and 
$\tau\tau$ constraints from $\gamma$-rays \cite{Cirelli:2012ut} in the green band and positrons \cite{Ibarra:2013zia} 
in the brown (dashed-dotted) line.
Finally to compare to more general decaying DM candidates, we also give in the black dashed line the exclusion limit obtained using 
the continuum $\gamma$-ray data from the Fermi-LAT and in the red solid line the exclusion limit derived using the AMS-02 positron flux 
observations in \cite{Cirelli:2012ut} and \cite{Ibarra:2013zia} respectively. These two analyses apply to generic decaying DM models where the 
DM candidate dominantly decays into a pair $\mu^{+}\mu^{-}$.  Also we show with the green (dotted) line projected limits from ICEcube on the
decay into two neutrinos~\cite{Covi:2009xn}. 
}
\label{tau}       
\end{figure}

The decay channels of gravitino DM depend on the particular realization of R-parity breaking. In the case of bilinear R-parity
breaking, the main decay channels are into a neutrino and a gauge boson, i.e. photon, Z or Higgs, or a charged lepton
and a $W$. The exact branching ratios depend on the gravitino mass and the supersymmetric spectrum. For a light gravitino below the 
$W$ threshold, the decay goes into a neutrino and photon, giving rise to the possibility of the smoking-gun signal of a photon line. 
For particular configurations of the gaugino masses or low $m_{3/2}$, the gravitino decay can be sufficiently suppressed 
to allow for R-parity breaking couplings able to generate also the neutrino masses~\cite{Restrepo:2011rj}.
In general though, the Fermi-LAT data set a strong bound on the DM lifetime in a photon line of the order of 
$ 5 \times 10^{28} $ s~\cite{Abdo:2010nc, Vertongen:2011mu}, excluding the R-parity breaking parameter space giving origin to
neutrino masses, if the gravitino is heavier than few GeV. The spectra from gravitino decay in bilinear R-parity  
models have been computed in~\cite{Bertone:2007aw, Ishiwata:2008cu, Covi:2008jy, Buchmuller:2009xv, Choi:2009ng}. 
In the case of trilinear R-parity breaking, also three-body decays can be important, because the 2-body decay only arises at one-loop 
level~\cite{Lola:2007rw,Bomark:2009zm,Bajc:2010qj}.
As for the case of DM annihilation, until now no clear signal for a DM decay has been found so far. Interpreting the
present data as a constraint in the case of gravitino DM, we obtain limits on its lifetime as shown in Fig.~\ref{tau}
collecting results on the bilinear R-parity breaking model from \cite{Vertongen:2011mu, Grefe:2011dp, Delahaye:2013yqa}
and adapting results from \cite{Ibarra:2013zia,Cirelli:2012ut}.
These gravitino lifetime limits can be reinterpreted in bilinear R-parity breaking models in a limit of the order $3 \times 10^{-8}$ 
on the bilinear R-parity breaking strength $\epsilon \sim \frac{\mu^{\prime i}}{\mu} $ below the $W\ell $ channel~\cite{Vertongen:2011mu}
and in even more stringent limits at larger gravitino masses~\cite{Delahaye:2013yqa}. 
For the axino case, since the couplings are less suppressed, the R-parity breaking interaction has instead to be of the order of
$ \sim 10^{-11}$ for axino masses around the GeV~\cite{Covi:2009pq}.
Regarding the claim of a tentative line at 130 GeV in the Fermi-LAT data~\cite{Bringmann:2012vr, Weniger:2012tx}, such a line signal may 
occur naturally with the correct intensity in bilinear R-parity breaking models, both from gravitino~\cite{Buchmuller:2012rc,Liew:2013mta} or axino \cite{Endo:2013si} DM decay,
but unfortunately the morphology of the signal region, which appears strongly concentrated in the galactic center \cite{Weniger:2012tx, Su:2012ft},
is not well-fitted by the more broadly distributed decaying DM signal~\cite{Buchmuller:2012rc}.
On the other hand, decaying gravitino DM can accommodate the positron excess observed by PAMELA~\cite{Buchmuller:2009xv} and AMS-02~\cite{Ibe:2013nka}.

\subsubsection{LHC searches}

For SuperWIMP DM the direct production at the LHC is in many cases too suppressed to allow to measure the
DM candidate directly. The single gravitino production rate at LHC has been computed in \cite{Klasen:2006kb}
and it results into a visible monojet signal only for very light gravitino with mass well below 1 eV. Such a gravitino 
could only be considered as a subdominant hot DM component, since the case of dominant hot DM is excluded by large scale
structure observations ~\cite{Viel:2007mv, Boyarsky:2008xj}.

So if we require the gravitino to be heavier than $\sim 100 $ keV~\footnote{This lower value for a Cold DM candidate 
is usually quoted for a classical thermal relic. In reality the boundary between Warm and Cold DM is fuzzy and depends strongly 
on the production mechanism, ensuing velocity distribution and possibly presence of other DM components, see e.g. ~\cite{ Boyarsky:2008xj}.}, 
the first evidence for gravitino (or axino) would be given by the observation of the strongly interacting superpartners like the gluino or
the squark, as already discussed in the neutralino case. The only difference is the modification of the decay chains, and the possibility to 
have the final decay of the NLSP into gravitino, axino or SM particles in the detector.

We can therefore distinguish two broad classes of signatures depending on the NLSP lifetime: either the NLSP decays in
the detector or it is stable on collider timescales. In general the NLSP decay can proceed either via the R-parity 
conserving (RPC) or the R-parity violating (RPV) couplings and the corresponding lifetimes, assuming e.g. a pure Bino NLSP, are
\begin{eqnarray}
\tau_{\chi, RPC, \tilde G} &=& 1.8 \times 10^{-3} \mbox{s} 
\left(\frac{M_1}{200~\rm GeV} \right)^{-5} 
\left(\frac{m_{3/2}}{1~\rm MeV}\right)^2 \nonumber\\
\tau_{\chi, RPC, \tilde a} &=& 3.1 \times 10^{-2}\mbox{s} 
\left(\frac{M_1}{200~\rm GeV} \right)^{-3} 
\left(\frac{f_a}{10^{11}~\rm GeV}\right)^2 \nonumber\\
\tau_{\chi, bilRPV} &\sim &  1.4 \times 10^{-6}\mbox{s} 
\left(\frac{M_1}{200~\rm GeV} \right)^{-1} 
\left( \frac{\epsilon}{10^{-9}} \right)^{-2} 
\end{eqnarray}
where we consider the RPC decays into photon and gravitino/axino, and the bilinear R-parity breaking decay into an electroweak 
gauge boson and a lepton, with $\epsilon $ denoting again the overall bilinear R-parity breaking coupling (for more details see e.g. \cite{Bobrovskyi:2011vx}).
We see from the time-scales involved that the neutralino decay is prompt (without resolvable second vertex, i.e. $\tau \leq 10^{-12}$ s) only for very 
small gravitino mass, very large Bino mass or large R-parity breaking couplings,  which are already excluded by indirect detection observations.
Therefore the classical collider analysis for gauge-mediated SUSY breaking models~\cite{Giudice:1998bp} or RPV models~\cite{Barbier:2004ez} 
with prompt vertices does not apply to the scenario of SuperWIMP DM. 

Much more probable is to have displaced vertices, as studied in \cite{Ishiwata:2008tp, Chang:2009sv, Meade:2010ji, Bobrovskyi:2011vx, Asai:2011wy},
and in that case, if the decay can be observed, the type of daughter particles will give information on the model and the presence of R-parity violation.
The phenomenology expected depends strongly on the type of NLSP and its decay channels. 
The signal from a long-lived neutralino (N)LSP, produced by squark and providing a muon in the final state, has been recently analyzed by the ATLAS 
collaboration~\cite{Aad:2012zx, ATLAS-CONF-2013-092} up to decay length of 1 m with no evidence of excess above the SM background.
Still many other possible NLSPs and decay channels are yet unexplored.

If instead the NLSP appears stable in the detector, then the most favorable case for detection is if the NLSP is charged.
In the case of a slepton NLSP, an electromagnetically charged track could be observed in the tracker and an escaping ``heavy muon'' in the 
muon chambers~\cite{Fairbairn:2006gg, Ellis:2006vu, Endo:2011uw, Ito:2011xs}, giving an unmistakable signal that a long decay and therefore a very weakly interacting sector
is present in the model.  Recent LHC analysis are given in \cite{Aad:2012pra, ATLAS-CONF-2013-058, Chatrchyan:2013oca} and they reach 
limits on the $\tilde \tau $ mass above 300 GeV for direct production.
Also colored metastable NLSP can give a rich signal, hadronizing into R-hadrons, that can also change
sign of the electric charge while they are moving in the detector \cite{Fairbairn:2006gg}.
The LHC collaborations are looking for such exotic metastable particles, and CMS sets already very strong constraints on 
collider-stable stop and gluino (N)LSP,  reaching a lower limits of the order of about  800 and 1200 GeV respectively \cite{Chatrchyan:2012sp, Chatrchyan:2013oca}. 
ATLAS is performing similar searches \cite{Aad:2012pra, ATLAS-CONF-2013-057}.

Note that a charged NLSP could also be captured in the detector or the surrounding material and open up the possibility to detect the
decay in the periods of no collider operation \cite{Asai:2009ka}, as long as the detectors are kept switched on. Moreover if one could
store a stau NLSP and measure not only the dominant decay, but also the radiative decay with an additional $\gamma $,
it would be possible to distinguish e.g. between gravitino or axino LSP~\cite{Brandenburg:2005he,Freitas:2011fx}.

For a detector-stable neutral NLSP, like the neutralino or the sneutrino, instead, the phenomenology is very similar to the classical supersymmetric WIMP 
scenario discussed in section \ref{wimpLHC}. 

\subsubsection{Complementarity of the different detection strategies}
If a DM signal will be seen in any of the above channels, it will be important to compare and check the
signatures also in an independent channel. In the case of gravitino DM with RPV, this may be possible
since the gravitino and NLSP decay derive from the same coupling and, especially for neutralino NLSP, they are
strongly correlated~\cite{Bobrovskyi:2011vx}. Therefore from the observation of a particular DM mass and lifetime in indirect detection,
one could infer at least a range of expected NLSP lifetimes and masses. On the other hand, measuring RPV at the LHC
would surely strongly restrict the possible DM candidates, and give a prediction of the possible SuperWIMP indirect
detection rate, depending on the DM mass.
In this scenario the direct detection measurement could instead be vital to disentangle the case of neutral NLSP
from the WIMP scenario, e.g. excluding the possibility that a neutralino seen at LHC could be the DM. 
 
Of course if the LHC will be able to measure the complete supersymmetric spectrum and estimate some of the neutralino
mixings, also detailed studies and checks on the possible DM production mechanisms could be realized extending
our knowledge of the cosmological history to the electroweak scale or beyond ~\cite{Choi:2007rh, Steffen:2008bt,Freitas:2009fb}.

\section{Dark matter and the Higgs boson}
\label{Higgs}
The discovery of a new bosonic state at the LHC whose mass is close 
to 126 GeV - plausibly (one of) the celebrated Higgs boson(s) -  
has influenced in various ways the last year of theoretical research 
in particle and astroparticle physics. Though the DM properties 
could be in principle unrelated to those of the Higgs boson, in the 
vast majority of the theoretical frameworks considered to quantitatively 
address the DM problem, the Higgs boson discovery indirectly 
impacts the allowed configurations in the parameter space of the 
underlying DM theory. The indirect influence of the Higgs boson 
mass on the nature of the DM candidate has been quantified by 
various groups through global fits of the most popular beyond the 
SM theories to large data sets including the latest LHC 
discovery and bounds obtained from the null result of searches for 
new physics~\cite{Strege:2012bt,Bechtle:2012zk,Buchmueller:2012hv}. 

In the context of the cMSSM the latest LHC Higgs boson mass measurement 
has significantly pushed towards larger values of $m_{1/2}$ the favored 
regions in the plane $(m_{1/2},m_0)$~\cite{Strege:2012bt}. This is mainly a consequence of the 
fact that global fits without the Higgs boson mass measurement tend to 
prefer Higgs boson masses below 120 GeV.  The requirement of reproducing 
the observed value for the Higgs mass therefore entails specific 
supersymmetric configurations characterized either by large stop masses,
enhancing the Higgs mass thorough sizable stop loop corrections, or by 
maximal mixing in the stop sector which occurs when the stop mixing 
parameter $X_t$ approaches the value 
$0.5(m_{\tilde{t}_1}^2 + m_{\tilde{t}_2}^2)$, where $m_{\tilde{t}_1}$ and 
$m_{\tilde{t}_2}$ are the masses of the stop mass eigenstates. 
In both scenarios low values of $m_{1/2}$ are not allowed. 

The impact of a Higgs boson mass close to 126 GeV has been also discussed 
in the NMSSM~\cite{Kowalska:2012gs} and in the context of the pMSSM~\cite{Arbey:2012bp}, where the large dimensionality 
of the underlying parameter space allows to accommodate more smoothly the 
latest LHC results. We refer the reader to the review of this series  
by John Ellis for further details on the role of the Higgs discovery in 
the context of SUSY global fits. We note only that in general a heavy spectrum is less easily reconciled
with the present DM density since the annihilation cross section determining
the WIMP relic density becomes weaker with increasing masses and therefore
the abundance larger. Nevertheless even in the simple cMSSM there are surviving
islands of acceptable neutralino number densities compatible with heavy
spectra, for example  in the Higgsino/Wino scenarios, which could give a stronger direct 
detection signal, or also along the stau-neutralino coannihilation strip.
In the latter case though, the degeneracy between the NLSP and LSP needed to
give the observed DM density is so strong that such region may be excluded soon by 
LHC searches for metastable massive charged particles~\cite{Citron:2012fg}.

In connection to the case of SuperWIMP DM, instead, a heavy
spectrum is not in general a problem, since it actually relaxes in part
the BBN bounds, thanks to the shorter NLSP lifetime. Moreover for large
NLSP masses it is easier to exploit the SuperWIMP production mechanism
also for charged NLSPs. On the other hand, one has to admit that a
heavy supersymmetric particle spectrum increases as well the thermal
production contribution to the present DM density tightening the upper bound on the 
reheat temperature~\cite{Hamaguchi:2009sz}.
Recent analysis of gravitino DM in different MSSM realizations after the Higgs discovery 
are given in Refs. \cite{Roszkowski:2012nq,Hasenkamp:2013opa}.

Though all LHC measurements 
are consistent with SM expectations, the observation of a Higgs diphoton 
rate somewhat larger than expected (the ATLAS collaboration reports a deviation close to 1.5 $\sigma$~\cite{Aad:2013wqa}), has motivated studies where 
this result was put in relation with the Fermi-LAT 130 GeV 
$\gamma$-ray line in models where DM is coupled to the Higgs boson. 
A simple unified description 
of the two phenomena has been proposed in Ref.~\cite{SchmidtHoberg:2012ip}, which provides a 
concrete realization within the NMSSM of a scenario where sizable DM 
annihilations into photons are associated with an enhanced Higgs diphoton rate.

\section{Summary and outlook}
\label{Summary}
We have reviewed here different supersymmetric DM candidates,
discussing in particular the two extreme cases of the neutralino WIMP
and the much more weakly interacting gravitino/axino.
In both cases we are still missing a convincing signal, though the LHC as well as 
direct and indirect detection experiments are already putting interesting constraints
on the parameter space of the basic supersymmetric models.
On the other hand, the supersymmetric framework for DM is still very flexible
and viable regions for (nearly) any type of supersymmetric DM are open
in more general settings and not only in those.

We would like to stress again that galactic and extragalactic DM searches provide
information on a different sector of the DM theory than that more directly
probed at the LHC and that data from all available sources will be
absolutely needed to identify univocally the DM particle.
In this respect we can look forward to a very productive time in the next decade 
since, while LHC will push the high energy exploration further, hopefully finding 
soon some type of beyond the SM physics, direct and indirect detection experiments 
will continue to search for DM. XENON 1-ton, for instance, 
is expected to start the scientific phase in 2015, essentially in parallel with the high-energy run of the LHC. 
The Cherenkov Telescope Array, an observatory for gro\-und-based $\gamma$-ray astronomy, is currently completing the preparatory phase while AMS-02 will continue taking high-quality data for many more years. Also the astrophysical aspects of the DM problem will be probed in depth in the next years through the Gaia space observatory, which will be launched in November 2013 to measure the kinematical properties of about one billion of stars in our galaxy.

In the best case scenario, there is still a reasonable hope of a contemporary
detection of DM in all three detection strategies and in many different
experiments, allowing to test thoroughly both the WIMP hypothesis
and the supersymmetric nature of DM. In absence of a signal at the LHC, direct and indirect detection experiments
may nevertheless be able to pinpoint a WIMP in the near future. On the other hand, for the case of
SuperWIMPs the LHC searches are not yet optimized, but
they will surely come to maturity in the next run, allowing to cover most
of the supersymmetric parameter space, as long as the mass scales
are within reach. In the worst case scenario, instead, where no DM signals will be identified 
in the next decade with the different techniques presented here, 
we would have nevertheless learned that many of our current paradigms, like for instance the WIMP-mechanism, need to be revised in favor of different and more flexible approaches, showing once again that Nature is often more rich and sophisticated than what one might expect at first. 

\section{Acknowledgements}

The authors acknowledge partial financial support by the German-Israeli
Foundation for scientific research and development (GIF) and by the
EU FP7 ITN Invisibles (Marie Curie Actions, PITN-GA-2011-289442).

 \bibliographystyle{iopart-num}
 \bibliography{references}
 

\end{document}